# Additive manufacturing for energy storage: Methods, designs and materials selection for customizable 3D printed batteries and supercapacitors

Umair Gulzar[1], Colm Glynn[2] and Colm O'Dwyer[1,3,4,5]*

[1]School of Chemistry, University College Cork, Cork, T12 YN60, Ireland
[2] Analog Devices International, Raheen, Limerick, Ireland
[3] Micro-Nano Systems Centre, Tyndall National Institute, Lee Maltings, Cork, T12 R5CP, Ireland
[4]AMBER@CRANN, Trinity College Dublin, Dublin 2, Ireland
[5]Environmental Research Institute, University College Cork, Lee Road, Cork T23 XE10, Ireland

**Abstract**

Additive manufacturing and 3D printing in particular have the potential to revolutionize existing fabrication processes where objects with complex structures and shapes can be built with multifunctional material systems. For electrochemical energy storage devices such as batteries and supercapacitors, 3D printing methods allows alternative form factors to be conceived based on the end use application need in mind at the design stage. Additively manufactured energy storage devices require active materials and composites that are printable and this is influenced by performance requirements and the basic electrochemistry. The interplay between electrochemical response, stability, material type, object complexity and end use application are key to realising 3D printing for electrochemical energy storage. Here, we summarise recent advances and highlight the important role of methods, designs and material selection for energy storage devices made by 3D printing, which is general to the majority of methods in use currently.

*Corresponding author: Email: c.odwyer@ucc.ie; Tel: +353 21 4902732

## 1. Introduction

The technology of additive manufacturing (AM), initially introduced in 1980s for building models and prototyping, is now commercially available in various forms of 3D printers. Contrary to conventional formative and subtractive manufacturing, the AM alias of 3D printing is capable of manufacturing high quality customizable parts from polymers, metals and ceramics without the expense of moulds or machining[1, 2]. Different methods of AM have been developed in last two decades which are classified by American Standards for Testing and Materials (ASTM) as (1) Material Jetting and (2) Extrusion (3) vat-photopolymerization (4) powder bed fusion (5) binder jetting (6) sheet lamination and (7)* direct energy deposition[3]. The capabilities and selection of each printing method and materials are detailed elsewhere[1, 4-9] and lies beyond the scope of this article. Though, the underlying principal of all AM methods involves the use of a computer aided design (CAD)-based virtual object for controlling the position of a material dispensing/building device. The object is constructed layer-by-layer, with layer thickness ranging from 15 to 500 μm, using building materials optimized for a specific printing method[10]. This way of fabrication allows direct manufacturing of final or near-final components with minimal post-processing, smaller operational foot print and maximum material utilization to achieve zero waste on-demand manufacturing[8].

AM is already well known in the field of medicine for making surgical guides and custom-made prosthetics[11, 12]. The tomographic data of a patient is used to produce a CAD design which can be 3D printed according to the size and shape of each individual patient. Recently, an integrated tissue-organ printing system has been developed to generate freeform shape with multiple types of cells and biomaterials[13]. Besides, various new materials including nanomaterials, functional/smart materials or even fast drying concrete, have been explored for 3D printability[14]. In fact, a Chinese company has already demonstrated the capability of AM by 3D printing multiple houses in a single day[9, 15]. All these technological achievements show that 3D printing has the potential to revolutionize the process of traditional manufacturing from aerospace to construction and electronic industry.

Electrochemical energy storage (EES) represents another important arena where unique building properties of AM and 3D printing can be exploited[16]**. Thoughtfully designed 3D structures are reported to show better performance in batteries and supercapacitors[17, 18]. Traditional EESDs construction include electrode fabrication, electrolyte addition and device assembly. Although these processes are well optimized for an assembly line production, 3D printed EESDs are desirables in markets with high demand for customization, flexibility and design complexity. Moreover, it can also provide the integration platform for EESDs and external electronics avoiding additional steps. Nevertheless, many technological challenges need to be addressed before realizing a complete 3D printed energy storage systems. This Opinion only explores the recent use of AM in the field of electrochemical energy storage devices (EESDs), mainly 3D printed batteries and supercapacitors. Moreover, different design strategies, printing methods and compatible materials already used in fabricating EESDs are discussed along with critical challenges and future prospects.

## 2. 3D printed electrochemical storage devices (EESDs)

Fabrication of an electrochemical energy storage device has its own challenges mainly due to hierarchical assembly of each individual component i.e. current collectors, electrodes, separators and electrolyte.

Moreover, each printing method has a unique way of printing an object using specific feed material. For example, powder bed fusion and binder jetting require solid feed while vat-photopolymerization, jetting and direct ink writing need a liquid feed. Hence, understanding the capabilities of each printing method, feed materials along with the overall design and chemistry of EESDs needs to be considered prior to the process of 3D printing.

*2.1. Materials and methods consideration*

Selecting materials and printing method is a synergetic process where materials are formulated according to the demands of the printing process based on the projected use in an EESD. As most of EESDs are fabricated through material jetting or extrusion (i.e. Inkjet printing (IJP), direct ink writing (DIW) and fused deposition modelling (FDM)), the characteristics of the raw material or ink are detrimental to mechanical and electrochemical performance of the final device. For example, low viscosity inks form better droplets (ideal for inkjet printing) while highly viscous inks tend to make continuous filaments that are suitable for extrusion based printing. For inkjet printing, viscosity ($\mu$), surface tension ($\sigma$), density ($\rho$) and the nozzle diameter (d) can be optimized using Ohnesorge number $Z = \sqrt{\rho \sigma d}/\mu$ and ink compositions with $1 < Z > 10$ are generally expected to produce stable droplets[10, 19, 20]. One of the problems generally faced during ink formulation is the aggregation of conductive agents (i.e. carbon, graphene, CNTs, metal nanoparticles) which clogs the extruding nozzle effecting the structure and the performance of EESD. Challenged by the same problem, Li et al.[21] used a systematic approach **(Figure 1a)** by first exfoliating 2D material (graphene and Molybdenum disulphide) into single or few-layers nanosheets in DMF ($\mu$ = 0.92 mPa·s) using a well-established liquid-phase exfoliation technique followed by the addition of a compatible polymer (cellulose) in order to reduce restacking of 2D materials. Later, another miscible solvent (terpineol; $\mu$ = 0.4 Pa·s) with lower toxicity and higher boiling point is added to the dispersion while the exfoliating solvent (DMF) is distilled off. Due to low viscosity requirement[22] for inkjet printing ($\mu$ = 0.1 Pa·s), the concentration of active material is limited to 2.0 g L$^{-1}$ and 0.12 g L$^{-1}$ for aqueous and organic dispersions, respectively[23, 24]. Hence, the dispersion (in terpineol) was tailored with a third solvent (ethanol) to achieve the required rheology for inkjet printing. The authors conclude that the selection of exfoliating solvent, stabilizing polymer, printable and tailoring solvent are all important for an optimal inkjet printing process.

Contrary to IJP, inks for DIW must exhibit shear thinning behaviour with high stress and storage modulus allowing shape retention of the extruded material during the process of deposition. Besides, the inks must have a rapid solidification process and mechanical stiffness to support subsequent layers. As the process uses highly viscous paste of feed material, not only the risk of clogging is reduced but high mass loading of active materials can be achieved allowing significant improvement in areal capacity of EESD. As an example, Kim et al.[25] formulated a polyvinylpyrolidone-wrapped multiwalled carbon nanotubes (PVP-MWCNT) based ink (7% MWCNTs and 17% PVP in water) with appropriate rheological properties for DIW **(Figure 1b)**. PVP was added to avoid agglomeration and nozzle clogging while the concentration MWCNTs was high as 75% inside the final 3D structure. With such high MWCNTs content, authors were able to achieve an electrical conductivity of 2540 S m$^{-1}$ and highlighted the potential of using highly conductive inks for 3D printed EESDs.

FDM is also an extrusion based printing method, contrary to DIW, it uses solid feed materials like acrylonitrile-butadiene-styrene (ABS) or polylactic acid (PLA) which are melted through a computer controlled nozzle and gets solidified immediately in a pre-defined 3D structure. Solidification of each layer is principally based on crystallization and chain entanglement of the polymer, however, addition of additive materials can affect the properties and solidification process of the matrix polymer. Common additives used in polymer matrix for EESDs are various conductive materials like ABS/graphene[26], ABS/carbon[27], PLA/graphene[28] and even PLA/LTO/carbon and PLA/LFP/carbon[29]* which are essential for electrode fabrication in lithium ion batteries. A good example is a recent study[30]** where three conductive agents (Super-P, MWCNTs, graphene) and two active materials (Lithium titanate, lithium manganese oxide) were blended with PLA to test the printability, conductivity and charge storage capacity of the new composite. Results showed that 30% of graphene, 20% of MWCNTs and 12% of Super-P can be mixed with PLA without compromising their printability while maximum storage capacity was obtained using 80:20 ratio of conductive and active material **(Figure 1c)**. Regarding extremely low capacities (1-2% of theoretical capacities) obtained for EESDs, authors claimed that it is the result of large % volume of PLA (70-80%) which was preventing conductive agents to make electrical contact with the active material. Therefore, continuous efforts are being made to increase the content of conductive agents without effecting the process of FDM based printing[27, 28, 31].

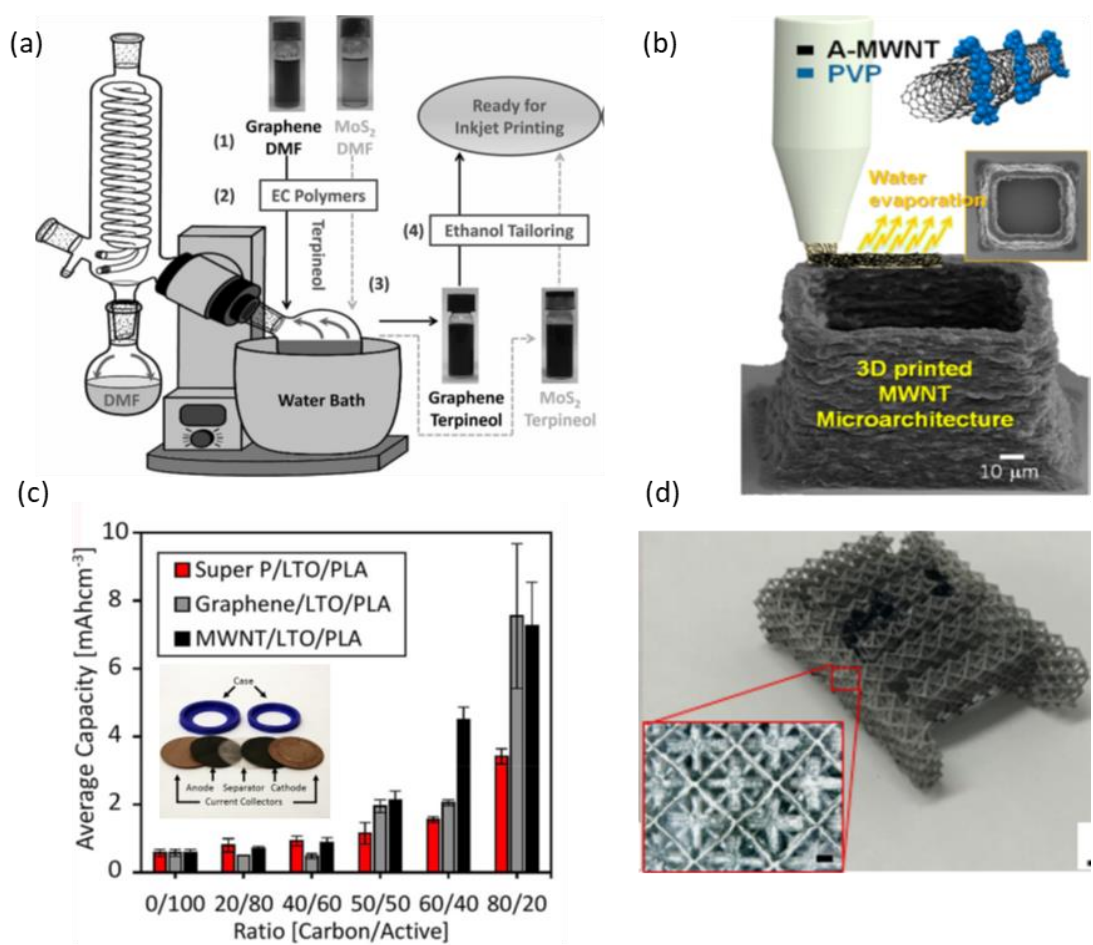

**Figure 1.** (a) Systematic strategy for preparing ink formulation for IJP. (b) DIW of PVP-MWCNTs based ink formulation, (c) Capacities and different composition of conductive additives for FDM printed EESD. (inset) FDM based printed electrodes and casing. (d) SLA-based printed EESD after pyrolysis. Reprinted with the permission of Wiley publications and American Chemical Society.

Moreover, it is known that impurities in lower grade poly(lactic acid) and ABS plastics can influence some electrochemical activity, and any method to increase the surface area of graphite-containing thermoplastic by solvent[32]* or thermal decomposition[33] will obviously increase material-electrolyte interactions. However, the intrinsic electronic and ionic conductivity in any printed electrode support material or active material composite is paramount, especially in the out-of-plane direction in sandwich design, and along the plane of in-plane designs. This is critical for intra- and inter-particle conductivity so that all active material in a composite, or active material are the surface is electrically and thus electrochemically addressable in a battery or supercapacitor. In the example in **Figure 2**, we show how the intrinsic surface of a graphite-loaded PLA composite formed by FDM printing, can be modified by either solvent or thermal decomposition, resulting in significant activation and enhancement of electrochemical activity for HER and OER reactions, and for 'switching on' reversible galvanostatic charging and discharging in the 3D printed electrodes of a printed PLA/ABS battery. In this case, we developed cell-to-cell clickable 3D printed battery cells using ABS outer casing and graphite-loaded PLA as the current collector. These PLA electrodes were coated which $LiMn_2O_4$ (anode) and $LiCoO_2$ (cathode) slurried with carbon nanotubes (**Figure 2a-c**), and separated by a $SiO_2$-PVP gel containing $LiNO_3$. This type of cell used acidic treatment of the PLA to open up the surface and significantly improve interfacial contact between the active deposited material and the graphitic conductive additive in the PLA. As this is an example of the sandwich type battery cell, the conductivity out of plane of the electrode was important. Prior to PLA surface decomposition, **Figure 2d** shows that the redox activity was negligible for this cell, and no reversible lithiation process found. Post activation (**Figure 2e**), the cell was able to charge and discharge efficiently, holding a capacity of ~80 mAh $g^{-1}$. This basic concept is fundamental to all 3D printing materials for electrochemical technologies, whether as supports, electrodes, current collectors or active material composites. Ionic and electronic conductivity need to be controlled and optimized where possible (**Figure 2f**), and these needs dictate both the materials choice and the method of printing from the outset.

Another possibility of printing highly complex 3D EESDs is to use vat-photopolymerization method which on one hand offers layer resolutions up-to 50 µm[34] but limited by the choice of materials. It uses a photo-curable resin consisting of monomers (acrylates or epoxy), photoinitiators (2,2-Dimethoxy-1,2-phenylacetophenone (DMPA)), diluents (1,6 hexanediol diacrylate), chain transfer agents (Allyl sulphides) and coupling agents. Monomers and photoinitiators are the main ingredients of the resin while the diluent, chain transfer and coupling agents are used to manipulate viscosity, degree of crosslinking and the bondage between reinforcement material and the resin, respectively[35]. Among them, the viscosity of the feedstock is an important parameter which can affect the quality and build speed of the object and generally optimized through diluents or controlling temperature. One recent study[36] provided a simple tool to investigate the relationship between photocurable properties of the resin containing viscosity increasing agents while another study showed how viscosity enhancing agent can be used to prevent particle agglomeration which resulted in a stable dispersion for more than 10 days[37].

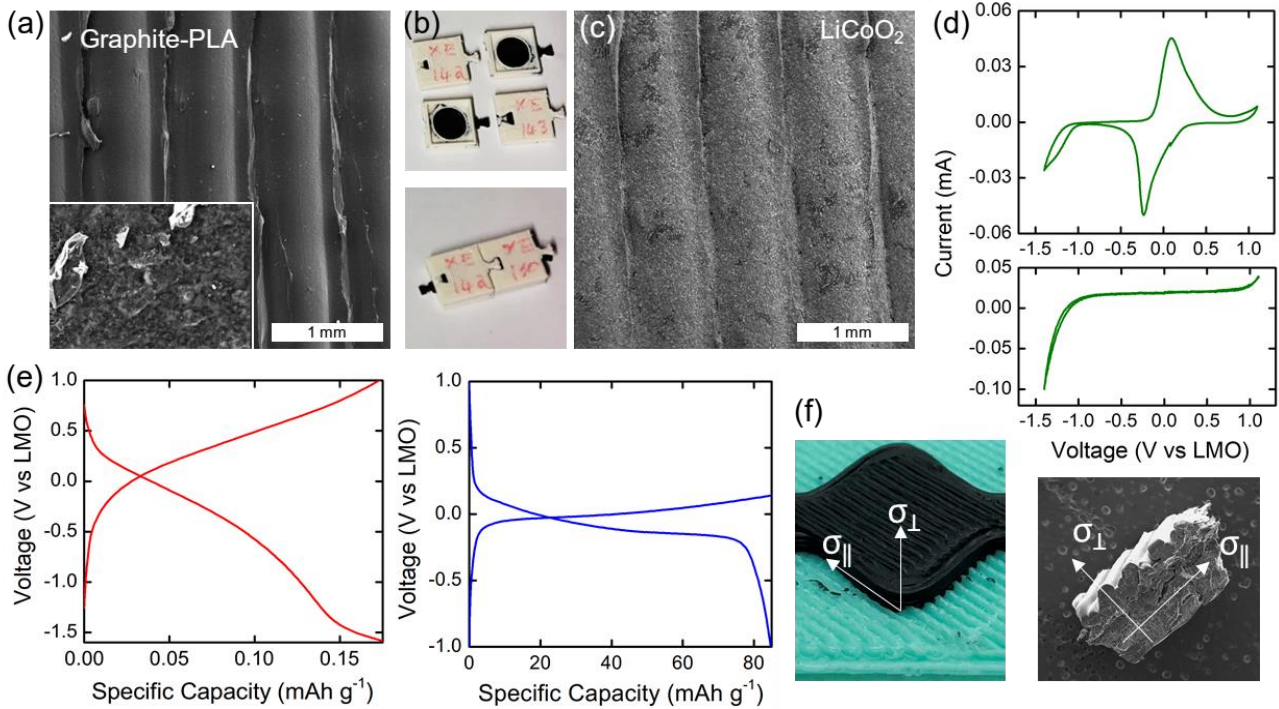

**Figure 2.** (a) SEM images of a graphite-PLA printed using FDM. Inset shows the porous morphology after acidic porosification of the surface. (b) Image of the click 3D printed batteries. (c) SEM of the LiCoO$_2$/CNT coated FDM printed PLA electrode (black region in (b)). (d) Cyclic voltammograms of the activated and as-printed graphite-PLA electrodes at scan rate of 0.5 mv s$^{-1}$ in a flooded solution of aqueous LiNO$_3$ electrolyte showing redox activity once PLA surface is porous. (c) Galvanostatic charge-discharge profiles of the as-printed and activated PLA electrodes in the full battery cell displaying over 2 orders of magnitude improvement in specific capacity post activation. (f) Section of FDM printed PLA on ABS (green) and corresponding SEM image of a PLA cross-section. σ$_\perp$ and σ$_\parallel$ represent perpendicular (out-of-plane) and parallel (in-plane) electrical conductivity.

Similar to FDM, SLA-based printing of EESDs require conductive agents for the fabrication of EESDs. One way is to deposit a metal layer after printing the desired shape[38] while second approach is to incorporate conductive agent inside photocurable resins. Addition of silver nitrate[39] and MWCNTs[40, 41]* in polyethylene glycol diacrylate (PEGDA) and acrylic based resin has already been reported with limited electrical conductivity. However, Park et al.[42] used silver nanowires as conductive fillers inside acrylate resin to construct a mechanically durable microstructure design. The conductivity of silver containing polymer structure still showed high resistance of 200 MΩ which was later reduced to 40 Ω using pyrolysis of the printed structure without compromising its structural integrity **(Figure 1d)**. Other methods like powder bed fusion, laminated object manufacturing and direct energy deposition are rarely been used to fabricate a complete EESD, nevertheless, some reports have shown manufacturing metal current collectors for EESDs[43-46]. Moreover, Senvol, a search engine for AM machines and related materials, provides useful information about the properties of more than thousand different materials used in all commercial AM machines[47].

### 2.1.1. Design considerations

Besides the geometrical architecture of individual electrodes, 3D printed batteries and supercapacitors are mostly assembled using an in-plane or sandwiched design **(Table 1)**. Each configuration has its own advantages and disadvantages, and also affect the electrochemical performance of EESDs and hence their application areas. For example, the sandwiched type EESDs are cost-effective with a potential of mass production. In-place designs allow minimum footprint with enhanced ionic transport making it suitable for tailored applications for ultrathin film batteries or supercapacitors. Exploring the potential of in-plane design, Sun et al.[48] investigated the effect of electrode thickness by printing multiple layers of electrode material and found that areal and volumetric capacitance of a supercapacitor show a linear increase with the number of printing layers. Similarly, Lin et al.[49] showed that the areal capacity of an in-plane capacitor with shorter interspace is higher than the capacitors with longer interspaces. However, printing compact designs with shorter interspaces and thicker electrode layers is challenging due to the rheological properties of the conductive ink which consists of binders, solvents, additives and active materials.

**Table 1.** 3D printed batteries and supercapacitors built using different designs and 3D printing methods.

| | Sandwiched Design | | | | | In-Plane Design | | | |
|---|---|---|---|---|---|---|---|---|---|
| | 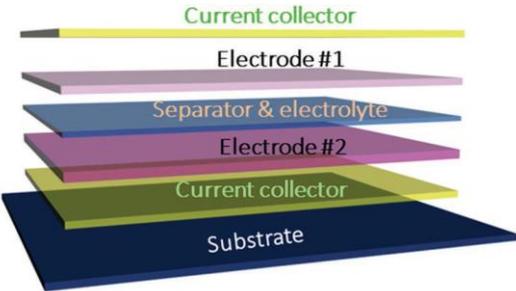 | | | | | 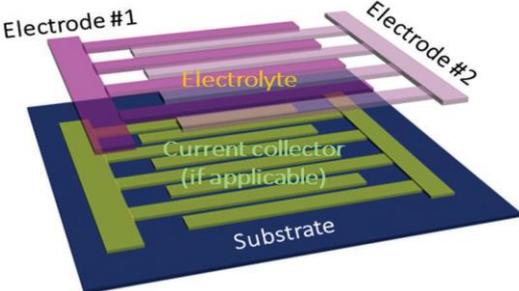 | | | |
| Type | Tech | Materials | Performance | Ref | Type | Tech | Materials | Performance | Ref |
| Batteries | IJP | Ag NPs | 5 mA h cm$^{-2}$ | [50] | Batteries | IJP | rGo/LTO/NCA | 0.35 mAh | [51] |
| | DIW | LTO/LFP/SP/PVP | 14.5 mA h cm$^{-2}$ | [52]** | | DIW | LTO/LFP/rGO | 1.5 mAh cm$^{-2}$ | [53, 54] |
| | FDM | LTO/LMO | 3.91 mAh cm$^{-3}$ | [30] | | FDM | -- | -- | -- |
| | SLA | LTO/LFP | 500 mAh cm$^{-2}$ | [55] | | SLA | NiSn/LMO | 2 µAh cm$^{-2}$ µm$^{-1}$ | [56] |
| Supercapacitors | IJP | PANI-GP | 864 F g$^{-1}$ | [57] | Supercapacitors | IJP | rGO | 0.1 mFcm$^{-2}$ | [58] |
| | DIW | | | | | DIW | PANI/GO | 1329 mFcm$^{-2}$ | [59] |
| | FDM | PLA/Graphene | 485 µF g$^{-1}$ | [28] | | FDM | ABS/CB | 12 µF cm$^{-2}$ | [27] |
| | SLA | Polymer/NiP/rGO | 250 mF cm$^{-2}$ | [38] | | SLA | Pyrolized polymer Ag NWs | 0.206 mF cm$^{-2}$ | [42] |

One observation worth noting during our literature search was that in-plane designs are preferred (or at least more common) when EESDs are fabricated using DIW and IJP, while sandwich designs is readily used for FDM and SLA-based 3D printing. We believe that FDM and SLA uses insulating polymer matrix and principles behind these printing methods allow limited choice for making conductive composites which are essential for constructing an EESD. A second more fundamental reason is that FDM and especially SLA-based 3D printing can create objects in full form factor directly (coin cells, thin film cells, outer casing as well as complex electrodes) in a single or multistep print. IJP by comparison is essentially a planar printing process whose 3D construction is a bottom up process and this fundamentally limits the complexity of single printed structures. In a recent study[53] SLA technology was used to make polymer graphene based conductive substrates, which were then electrophoretically coated with anode material (LTO), LiAlO$_2$-PEO membrane and cathode material (LFP) using a tri-layer sandwiched design. Cells cycled at 0.1C provided areal capacity of 400-500 µAh cm$^{-2}$.

Whether its sandwiched or in-plane design, the nature of charge storage process is always an important consideration before the selection of materials and method. Surface pseudocapacitive storage and electrochemical double layer capacitance will benefit from a higher surface to volume ratio, which can be achieved by etching or selective decomposition of the composite printed thermoplastic or photocurable resin. This is useful so long as the mechanical integrity of the printed object is not comprised by solvent, heating or excessive degree of porosity. Secondly, a high surface area material capable of capacitive charge storage is only useful in devices if the printed material is sufficiently electrically conductive, and mass loading within the feed material prior to printing is considered. Thus, the 3D printing technique involves final device operation and material selection. However, the situation is more pronounced when the internal volume fraction of additive material must be both conductive and accessible to Li (or other) ions to maximize volumetric and gravimetric energy density or areal capacity.

## 3. Conclusions

With new developments and reducing cost, AM and 3D printing in particular has the potential to revolutionize existing fabrication process where objects with complex structures and shapes can be built with multifunctional material systems. Nevertheless, AM suffers from many challenges and most of these challenges[61, 62]* are centred around build speed, mechanical properties of the final product, resolution of each printed layer, potential of using conductive feed material with an option of multi-material 3D printing. For EESD applications, and to some extent other electrochemical system such as water splitting, hydrogen generation, photo-electrochemistry and electrochemical sensors as pertinent examples, the choice of material, the nature of the final print in terms of composition, together with the attributes of the print specific to the application, will dictate the printing method used. There remains much to be developed to make any truly 3D printable EESD, i.e. with a customizable, non planar shape or form-factor, competitive with most forms of existing li-ion and supercapacitor technology. Although, IJP as an approach for supercapacitors is making some headway in this regards, as the method and the aqueous electrolyte requirements are less stringent that those for higher voltage Li-ion or alternative battery chemistry systems.

**Acknowledgements**

This work is supported by European Union's Horizon 2020 research and innovation programme under grant agreement No 825114. We also acknowledge funding support from Science Foundation Ireland (SFI) under Awards no. 14/IA/2581 and 15/TIDA/2893, and from the Irish Research Council Advanced Laureate Award under grant no. IRCLA/2019/118. The authors thank Dr Vladimir Egorov for FDM printing of the PLA-ABS sample.

**Declaration of interest**

None

## References

[1] Z. Chen, Z. Li, J. Li, C.C. Liu, C. Lao, Y. Fu, C.C. Liu, Y. Li, P. Wang, Y. He, 3D printing of ceramics: A review, Journal of the European Ceramic Society 39(4) (2019) 661-687.

[2] T.D. Ngo, A. Kashani, G. Imbalzano, K.T.Q. Nguyen, D. Hui, Additive manufacturing (3D printing): A review of materials, methods, applications and challenges, Composites Part B: Engineering 143 (2018) 172-196.
[3] Astm, Standard Terminology for Additive Manufacturing – General Principles – Terminology, West Conshohocken PA, 2016.
[4] A. Bandyopadhyay, B. Heer, Additive manufacturing of multi-material structures, Materials Science and Engineering: R: Reports 129 (2018) 1-16.
[5] J.R.C. Dizon, A.H. Espera, Q. Chen, R.C. Advincula, Mechanical characterization of 3D-printed polymers, Additive Manufacturing 20 (2018) 44-67.
[6] S. Ford, M. Despeisse, Additive manufacturing and sustainability: an exploratory study of the advantages and challenges, Journal of Cleaner Production 137 (2016) 1573-1587.
[7] R. Jiang, R. Kleer, F.T. Piller, Predicting the future of additive manufacturing: A Delphi study on economic and societal implications of 3D printing for 2030, Technological Forecasting and Social Change 117 (2017) 84-97.
[8] S.A.M. Tofail, E.P. Koumoulos, A. Bandyopadhyay, S. Bose, L. O'Donoghue, C. Charitidis, Additive manufacturing: scientific and technological challenges, market uptake and opportunities, Materials Today 21(1) (2018) 22-37.
[9] P. Wu, J. Wang, X. Wang, A critical review of the use of 3-D printing in the construction industry, Automation in Construction 68 (2016) 21-31.
[10] S.C. Ligon, R. Liska, J. Stampfl, M. Gurr, R. Mülhaupt, Polymers for 3D Printing and Customized Additive Manufacturing, Chemical Reviews 117(15) (2017) 10212-10290.
[11] S.L. Sing, J. An, W.Y. Yeong, F.E. Wiria, Laser and electron-beam powder-bed additive manufacturing of metallic implants: A review on processes, materials and designs, (1554-527X (Electronic)).
[12] M. Walker, S. Humphries, 3D Printing: Applications in evolution and ecology, Ecol Evol 9(7) (2019) 4289-4301.
[13] H.-W. Kang, S.J. Lee, I.K. Ko, C. Kengla, J.J. Yoo, A. Atala, A 3D bioprinting system to produce human-scale tissue constructs with structural integrity, Nature Biotechnology 34 (2016) 312.
[14] C. Zhu, T. Liu, F. Qian, W. Chen, S. Chandrasekaran, B. Yao, Y. Song, E.B. Duoss, J.D. Kuntz, C.M. Spadaccini, M.A. Worsley, Y. Li, 3D printed functional nanomaterials for electrochemical energy storage, Nano Today 15 (2017) 107-120.
[15] Kira, WinSun China builds world's first 3D printed villa and tallest 3D printed apartment building, 2015. (Accessed 28/10/2019 2019).
[16] C.-Y. Lee, A.C. Taylor, A. Nattestad, S. Beirne, G.G. Wallace, 3D Printing for Electrocatalytic Applications, Joule 3(8) (2019) 1835-1849.
[17] D.R. Rolison, J.W. Long, J.C. Lytle, A.E. Fischer, C.P. Rhodes, T.M. McEvoy, M.E. Bourg, A.M. Lubers, Multifunctional 3D nanoarchitectures for energy storage and conversion, Chemical Society Reviews 38(1) (2009) 226-252.
[18] M. Osiak, H. Geaney, E. Armstrong, C. O'Dwyer, Structuring Materials for Lithium-ion Batteries: Advancements in Nanomaterial Structure, Composition, and Defined Assembly on Cell Performance, J. Mater. Chem. A 2(25) (2014) 9433-9460.
[19] X. Wang, M. Jiang, Z. Zhou, J. Gou, D. Hui, 3D printing of polymer matrix composites: A review and prospective, Composites Part B: Engineering 110 (2017) 442-458.
[20] F. Zhang, M. Wei, V.V. Viswanathan, B. Swart, Y. Shao, G. Wu, C. Zhou, 3D printing technologies for electrochemical energy storage, Nano Energy 40(May) (2017) 418-431.
[21] J. Li, M.C. Lemme, M. Ostling, Inkjet printing of 2D layered materials, Chemphyschem 15(16) (2014) 3427-34.
[22] T. Nathan-Walleser, I.-M. Lazar, M. Fabritius, F.J. Tölle, Q. Xia, B. Bruchmann, S.S. Venkataraman, M.G. Schwab, R. Mülhaupt, 3D Micro-Extrusion of Graphene-based Active Electrodes: Towards High-Rate AC Line Filtering Performance Electrochemical Capacitors, Advanced Functional Materials 24(29) (2014) 4706-4716.
[23] V. Dua, S.P. Surwade, S. Ammu, S.R. Agnihotra, S. Jain, K.E. Roberts, S. Park, R.S. Ruoff, S.K. Manohar, All-Organic Vapor Sensor Using Inkjet-Printed Reduced Graphene Oxide, Angewandte Chemie International Edition 49(12) (2010) 2154-2157.
[24] U. Khan, M. O'Neill A Fau - Lotya, S. Lotya M Fau - De, J.N. De S Fau - Coleman, J.N. Coleman, High-concentration solvent exfoliation of graphene, (1613-6829 (Electronic)).
[25] J.H. Kim, S. Lee, M. Wajahat, H. Jeong, W.S. Chang, H.J. Jeong, J.R. Yang, J.T. Kim, S.K. Seol, Three-Dimensional Printing of Highly Conductive Carbon Nanotube Microarchitectures with Fluid Ink, ACS Nano 10(9) (2016) 8879-87.
[26] X. Wei, D. Li, W. Jiang, Z. Gu, X. Wang, Z. Zhang, Z. Sun, 3D Printable Graphene Composite, Sci Rep 5 (2015) 11181.

[27] H.H. Bin Hamzah, O. Keattch, D. Covill, B.A. Patel, The effects of printing orientation on the electrochemical behaviour of 3D printed acrylonitrile butadiene styrene (ABS)/carbon black electrodes, Sci Rep 8(1) (2018) 9135.
[28] C.W. Foster, M.P. Down, Y. Zhang, X. Ji, S.J. Rowley-Neale, G.C. Smith, P.J. Kelly, C.E. Banks, 3D Printed Graphene Based Energy Storage Devices, Sci Rep 7 (2017) 42233.
[29] H. Ragones, S. Menkin, Y. Kamir, A. Gladkikh, T. Mukra, G. Kosa, D. Golodnitsky, Towards smart free form-factor 3D printable batteries, Sustainable Energy & Fuels 2(7) (2018) 1542-1549.
[30] C. Reyes, R. Somogyi, S. Niu, M.A. Cruz, F. Yang, M.J. Catenacci, C.P. Rhodes, B.J. Wiley, Three-Dimensional Printing of a Complete Lithium Ion Battery with Fused Filament Fabrication, ACS Applied Energy Materials 1(10) (2018) 5268-5279.
[31] A. Maurel, M. Courty, B. Fleutot, H. Tortajada, K. Prashantha, M. Armand, S. Grugeon, S. Panier, L. Dupont, Highly Loaded Graphite-Polylactic Acid Composite-Based Filaments for Lithium-Ion Battery Three-Dimensional Printing, Chemistry of Materials 30(21) (2018) 7484-7493.
[32] M.P. Browne, F. Novotný, Z. Sofer, M. Pumera, 3D Printed Graphene Electrodes' Electrochemical Activation, ACS Applied Materials & Interfaces 10(46) (2018) 40294-40301.
[33] F. Novotný, V. Urbanová, J. Plutnar, M. Pumera, Preserving Fine Structure Details and Dramatically Enhancing Electron Transfer Rates in Graphene 3D-Printed Electrodes via Thermal Annealing: Toward Nitroaromatic Explosives Sensing, ACS Applied Materials & Interfaces 11(38) (2019) 35371-35375.
[34] J.Z. Manapat, Q. Chen, P. Ye, R.C. Advincula, 3D Printing of Polymer Nanocomposites via Stereolithography, Macromolecular Materials and Engineering 302(9) (2017) 1-13.
[35] A. Medellin, W. Du, G. Miao, J. Zou, Z. Pei, C. Ma, Vat Photopolymerization3D Printing of Nanocomposites: A Literature Review, Journal of Micro and Nano-Manufacturing 7(3) (2019).
[36] M. Yasui, K. Ikuta, Modeling and measurement of curing properties of photocurable polymer containing magnetic particles and microcapsules, Microsystems & Nanoengineering 3(1) (2017) 17035.
[37] K. Kobayashi, K. Ikuta, Three-dimensional magnetic microstructures fabricated by microstereolithography, Applied Physics Letters 92(26) (2008) 262505.
[38] J. Xue, L. Gao, X. Hu, K. Cao, W. Zhou, W. Wang, Y. Lu, Stereolithographic 3D Printing-Based Hierarchically Cellular Lattices for High-Performance Quasi-Solid Supercapacitor, Nano-Micro Letters 11(1) (2019) 46-46.
[39] E. Fantino, A. Chiappone, I. Roppolo, D. Manfredi, R. Bongiovanni, C.F. Pirri, F. Calignano, 3D Printing of Conductive Complex Structures with In Situ Generation of Silver Nanoparticles, Adv Mater 28(19) (2016) 3712-7.
[40] Q. Mu, L. Wang, C.K. Dunn, X. Kuang, F. Duan, Z. Zhang, H.J. Qi, T. Wang, Digital light processing 3D printing of conductive complex structures, Additive Manufacturing 18 (2017) 74-83.
[41] G. Gonzalez, A. Chiappone, I. Roppolo, E. Fantino, V. Bertana, F. Perrucci, L. Scaltrito, F. Pirri, M. Sangermano, Development of 3D printable formulations containing CNT with enhanced electrical properties, Polymer 109 (2017) 246-253.
[42] S.H. Park, M. Kaur, D. Yun, W.S. Kim, Hierarchically Designed Electron Paths in 3D Printed Energy Storage Devices, Langmuir 34(37) (2018) 10897-10904.
[43] C. Zhao, C. Wang, R. Gorkin, S. Beirne, K. Shu, G.G. Wallace, Three dimensional (3D) printed electrodes for interdigitated supercapacitors, Electrochemistry Communications 41 (2014) 20-23.
[44] X. Liu, R. Jervis, R.C. Maher, I.J. Villar-Garcia, M. Naylor-Marlow, P.R. Shearing, M. Ouyang, L. Cohen, N.P. Brandon, B. Wu, 3D-Printed Structural Pseudocapacitors, Advanced Materials Technologies 1(9) (2016).
[45] D.X. Luong, A.K. Subramanian, G.A.L. Silva, J. Yoon, S. Cofer, K. Yang, P.S. Owuor, T. Wang, Z. Wang, J. Lou, P.M. Ajayan, J.M. Tour, Laminated Object Manufacturing of 3D-Printed Laser-Induced Graphene Foams, Adv Mater 30(28) (2018) e1707416.
[46] M. Krinitcyn, Z. Fu, J. Harris, K. Kostikov, G.A. Pribytkov, P. Greil, N. Travitzky, Laminated Object Manufacturing of in-situ synthesized MAX-phase composites, Ceramics International 43(12) (2017) 9241-9245.
[47] Senvol, Senvol Database: Industrial additive manufacturing machines and materials. (Accessed 29/10/2019 29/10/2019).
[48] G. Sun, J. An, C.K. Chua, H. Pang, J. Zhang, P. Chen, Layer-by-layer printing of laminated graphene-based interdigitated microelectrodes for flexible planar micro-supercapacitors, Electrochemistry Communications 51 (2015) 33-36.
[49] A. Sumboja, M. Lübke, Y. Wang, T. An, Y. Zong, Z. Liu, All-Solid-State, Foldable, and Rechargeable Zn-Air Batteries Based on Manganese Oxide Grown on Graphene-Coated Carbon Cloth Air Cathode, Advanced Energy Materials 7(20) (2017) 1700927.

[50] M.S. Saleh, J. Li, J. Park, R. Panat, 3D printed hierarchically-porous microlattice electrode materials for exceptionally high specific capacity and areal capacity lithium ion batteries, Additive Manufacturing 23 (2018) 70-78.
[51] R. Hahn, M. Ferch, K. Tribowski, N.A. Kyeremateng, K. Hoeppner, K. Marquardt, K.D. Lang, W. Bock, High-throughput battery materials testing based on test cell arrays and dispense/jet printed electrodes, Microsystem Technologies 25(4) (2019) 1137-1149.
[52] T.S. Wei, B.Y. Ahn, J. Grotto, J.A. Lewis, 3D Printing of Customized Li-Ion Batteries with Thick Electrodes, Advanced Materials 30(16) (2018) 1-7.
[53] K. Fu, Y. Wang, C. Yan, Y. Yao, Y. Chen, J. Dai, S. Lacey, Y. Wang, J. Wan, T. Li, Z. Wang, Y. Xu, L. Hu, Graphene Oxide-Based Electrode Inks for 3D-Printed Lithium-Ion Batteries, Advanced Materials 28(13) (2016) 2587-2594.
[54] K. Sun, T.S. Wei, B.Y. Ahn, J.Y. Seo, S.J. Dillon, J.A. Lewis, 3D printing of interdigitated Li-ion microbattery architectures, Advanced Materials 25(33) (2013) 4539-4543.
[55] E. Cohen, S. Menkin, M. Lifshits, Y. Kamir, A. Gladkich, G. Kosa, D. Golodnitsky, Novel rechargeable 3D-Microbatteries on 3D-printed-polymer substrates: Feasibility study, Electrochimica Acta 265 (2018) 690-701.
[56] H. Ning, J.H. Pikul, R. Zhang, X. Li, S. Xu, J. Wang, J.A. Rogers, W.P. King, P.V. Braun, Holographic patterning of high-performance on-chip 3D lithium-ion microbatteries, Proc Natl Acad Sci U S A 112(21) (2015) 6573-8.
[57] K. Chi, Z. Zhang, J. Xi, Y. Huang, F. Xiao, S. Wang, Y. Liu, Freestanding Graphene Paper Supported Three-Dimensional Porous Graphene–Polyaniline Nanocomposite Synthesized by Inkjet Printing and in Flexible All-Solid-State Supercapacitor, ACS Applied Materials & Interfaces 6(18) (2014) 16312-16319.
[58] J. Li, V. Mishukova, M. Östling, All-solid-state micro-supercapacitors based on inkjet printed graphene electrodes, Applied Physics Letters 109(12) (2016).
[59] Z. Wang, Q.e. Zhang, S. Long, Y. Luo, P. Yu, Z. Tan, J. Bai, B. Qu, Y. Yang, J. Shi, H. Zhou, Z.Y. Xiao, W. Hong, H. Bai, Three-Dimensional Printing of Polyaniline/Reduced Graphene Oxide Composite for High-Performance Planar Supercapacitor, ACS Applied Materials and Interfaces 10(12) (2018) 10437-10444.
[60] T. Sang Tran, N.K. Dutta, N. Roy Choudhury, Graphene-Based Inks for Printing of Planar Micro-Supercapacitors: A Review, Materials (Basel) 12(6) (2019).
[61] P. Chang, H. Mei, S. Zhou, K.G. Dassios, L. Cheng, 3D printed electrochemical energy storage devices, Journal of Materials Chemistry A 7(9) (2019) 4230-4258.
[62] X. Tian, J. Jin, S. Yuan, C.K. Chua, S.B. Tor, K. Zhou, Emerging 3D-Printed Electrochemical Energy Storage Devices: A Critical Review, Advanced Energy Materials 7(17) (2017) 1-17.